\documentclass[dvips,12pt]{emulateapj}
\usepackage[english]{babel}
\usepackage{amsmath}
\usepackage{amssymb}
\usepackage{url}
\usepackage[dvips]{color}
\usepackage{natbib}

\newcommand{\ergs}{\rm\,erg\,s^{-1}}

\newcommand{\msun}{\,M_{\odot}}

\begin{document}

\title{A New Accretion Disk Around the Missing Link Binary System PSR J1023+0038}
\shorttitle{Accretion Disk Around PSR J1023+0038}
\shortauthors{Patruno et~al.}

\author{A. Patruno\altaffilmark{1,2}, A.~M. Archibald\altaffilmark{2}, J.~W.~T. Hessels\altaffilmark{2,3}, S. Bogdanov\altaffilmark{4}, B.~W. Stappers\altaffilmark{5}, C.~G. Bassa\altaffilmark{2}, G.~H. Janssen\altaffilmark{2}, V.~M. Kaspi\altaffilmark{6}, S. Tendulkar\altaffilmark{7}, A.~G. Lyne\altaffilmark{5}}
\altaffiltext{1}{Leiden Observatory, Leiden University, PO Box 9513, NL-2300 RA Leiden, the Netherlands}
\altaffiltext{2}{ASTRON, the Netherlands Institute for Radio Astronomy, Postbus 2, 7990 AA Dwingeloo, the Netherlands}
\altaffiltext{3}{Astronomical Institute A.Pannekoek, University of Amsterdam, 1098XH, Amsterdam, the Netherlands}
\altaffiltext{4}{Columbia Astrophysics Laboratory, Columbia University, 550 West 120th Street, New York, NY 10027, USA}
\altaffiltext{5}{Jodrell Bank Centre for Astrophysics, School of Physics and Astronomy, The University of Manchester, Manchester M13 9PL, UK}
\altaffiltext{6}{McGill University, 3600 University Street, Montreal, QC H3A 2T8, Canada}
\altaffiltext{7}{Space Radiation Laboratory, California Institute of Technology, 1200 E California Blvd, MC 249-17, Pasadena, CA 91125, USA}

\begin{abstract}

PSR J1023+0038 is an exceptional system for understanding how slowly
rotating neutron stars are spun up to millisecond rotational periods
through accretion from a companion star. Observed as a radio pulsar
from $2007-2013$, optical data showed that the system had an accretion
disk in 2000/2001.  Starting at the end of 2013 June, the radio pulsar has
become undetectable, suggesting a return to the previous
accretion-disk state, where the system more closely resembles an X-ray
binary. In this Letter we report the first targeted X-ray observations ever
performed of the active phase and complement them with UV/Optical and
radio observations collected in 2013 October. We find strong evidence
that indeed an accretion disk has recently formed in the system and we
report the detection of fast X-ray changes spanning about two
orders of magnitude in luminosity. No radio pulsations are seen during 
low flux states in the X-ray light-curve or at any other times.

\end{abstract}



\section{Introduction}

PSR J1023+0038 (henceforth J1023) is a 1.7-ms radio pulsar in a 4.8-hr
orbit with a ${\sim}0.2 M_{\odot}$ companion star \citep{arc09,
  arc10}. Before its 2007 discovery as a radio millisecond pulsar (MSP),
the source was known as a candidate quiescent low mass X-ray binary
(LMXB; \citealt{tho05, hom06}).  Strong evidence for the presence of
an accretion disk came from optical observations in $2000-2001$
(\citealt{tho05, bon02, wan09b}; see also \citealt{arc09} for an
overview).  Detailed optical/X-ray observations from 2002 onward suggest the absence
of an accretion disk \citep{arc09, arc10, bog11}. Therefore, it is now
accepted that, at least in 2001, J1023 was an accreting neutron star
in an LMXB; the neutron star (NS) has subsequently turned on as a radio MSP between
$2001-2007$.  From 2007 onwards this source has been consistently
observed as an eclipsing radio MSP (a so-called ``redback''
system, see \citealt{rob11}), with a low X-ray luminosity thought to
be largely due to a pulsar-wind-driven shock near the inner Lagrangian
point $L_1$ \citep{bog11}, probably with a contribution from the
pulsar's surface or magnetosphere \citep{arc10}.

According to the current paradigm, MSPs are formed in
X-ray binaries by accreting gas from a low-mass donor companion.  The
discovery of accreting millisecond X-ray pulsars (AMXPs; e.g.,
\citealt{wij98}) represented a first confirmation of this
``recycling'' scenario (see \citealt{pat12r} for a review). 
The accreted matter spins up the NS to millisecond periods. How and
when the accretion process switches off remains unknown (see
e.g.,~\citealt{tau12}).  J1023 has been heralded as the ``missing
link'' between radio MSPs and LMXBs and has given new insights into
this evolutionary process.

The very recent discovery of an X-ray active phase from PSR
J1824$-$2452I (a.k.a. IGR J18245$-$2452; spin period
3.9\,ms, orbital period 11\,hr; hereafter M28I, since it is in the
globular cluster M28) has again confirmed the recycling
scenario. Originally discovered in 2006 as a radio MSP, M28I was
observed to turn into a typical AMXP --- with active accretion onto
the NS surface and associated X-ray bursts --- and then return
surprisingly rapidly to an X-ray quiescent/radio MSP state within only
a few weeks after the X-ray outburst \citep{pap13a, pap13b}.  There
are several similarities between the J1023 and M28I
systems, but the former enjoys the advantage of being at a quarter the
distance and hosting a pulsar that is more than an
order-of-magnitude brighter at radio wavelengths.

In the period 2013 June 15-23, J1023 switched off as a radio pulsar~\citep{sta13}.
Solar constraints prevented the observation of this source with most X-ray
and UV/optical telescopes until mid-October. The
\textit{Swift}/Burst-Alert-Telescope was able to monitor the source in
hard X-rays, showing non-detections throughout the monitoring with
typical upper limits of $10^{35}\ergs$ in the $15-50$\,keV band (using
a distance of 1.368\,kpc, \citealt{del12}).

 In this Letter we report the first X-ray and UV/optical results on
 the radio-quiet phase of J1023 performed on 2013 October $18-19$ by
 the \textit{Swift} satellite.  The observations show increased X-ray
 and UV/optical activities (\citealt{kon13}, \citealt{pat13b}) that
 provide compelling evidence for the presence of a recently formed
 accretion disk in the system. We also report radio observations
 carried out with the Lovell and Westerbork Radio Synthesis Telescope
 that show no evidence for pulsed radio emission.

\section{Observations and Data Reduction}

We analyzed three targeted \textit{Swift} observations taken on 2013
June 10, 12 and October 18/19. This last observation started with two
short exposures on October 18 at UT 05:11:50 and was continued with six
more exposures on October 19 starting at UT 00:14:26. We used all data
collected with the \text{Swift}/X-ray-Telescope (XRT) and the
Ultra-Violet/Optical Telescope (UVOT). The XRT was operated in
photon-counting mode (2.5-s time resolution) in all three
observations, whereas the UVOT operated with the filters UW2
(June 10), UW1 (June 12) and UW1/U (October 18/19).  The total
on-source time was $\simeq14$\,ks, with 10\,ks on October 18/19 and
${\sim}2$\,ks on each of June 10 and 12.

We reduced the data using the UVOT and XRT pipelines and applying
standard event screening criteria. The October $18-19$ UVOT
observation started with the first two exposures taken with the UW1
filter and the remaining six with the U filter. We extracted source
events for each observation using circular regions with radii of
$40^{\prime \prime}$ (for XRT) and $8^{\prime \prime}$ (for UVOT) and
by using the best astrometric position available
\citep{del12}. We kept only X-ray photons
with energies between $0.5-10$\,keV and then removed the background by
measuring it in a $40^{\prime \prime}$ region far from
bright sources. We combined the UVOT exposures within each
observation to provide a single high-signal-to-noise image for each
filter. We also analyzed the six October 19 U-band exposures
separately to highlight the rapid time variability of the source. We
performed a power-spectral analysis of the $0.5-10$\,keV XRT data by
creating Fourier transforms of different length, between ${\sim}130$
seconds (64 data points) and ${\sim}1300$ seconds (512 data points)
all with a Nyquist frequency of $\simeq 0.2$\,Hz.

To complement these Swift observations, we also performed targeted
observations with the Lovell and Westerbork radio telescopes.  Lovell
observations began on 2013 Oct 19 at 07:32:44 UT and were obtained
using the ROACH backend, coherently dedispersing 400\,MHz of bandwidth
centered at a frequency of 5.1\,GHz.  The observations lasted 2\,hr
and spanned orbital phases 0.35--0.77, thus occurring away from the
usual eclipse time at this high radio frequency. A total of five
15--60-min Westerbork observations (three of which were at orbital
phases where we do not expect radio eclipses) were acquired on October
16, 20, and 21 at central frequencies of 350 and 1380\,MHz and with
respective bandwidths of 80 and 160\,MHz, using the tied-array mode
and PuMaII backend~\citep{kar08}.  Data from both telescopes were
folded using a recent rotational ephemeris and inspected on a variety
of timescales to look for pulsed radio emission. 

\section{Results}

The \textit{Swift} June 10/12 observations, occurring before the
reported disappearance of the radio MSP, show a dim X-ray source
compatible with the radio-loud state reported in \citet{hom06},
\citet{arc10},~\citet{tam10} and \citet{bog11}.  An inspection of the
light-curve shows a faint source with $0.5-10$\,keV count rate of
${\sim}0.01$\,ct/s and little variation throughout the
observations. The UVOT images also show marginal detections (${\sim}
20$\,mag) and non-detections ($>20.6$\,mag) in the UW1 and UW2
filters, respectively.  In the following we focus on the October 18/19
\textit{Swift} observations and on the radio observations carried
out after the disappearance of the radio MSP.

\begin{figure*}[t!]
  \begin{center}
    \rotatebox{0}{\includegraphics[width=1.0\textwidth]{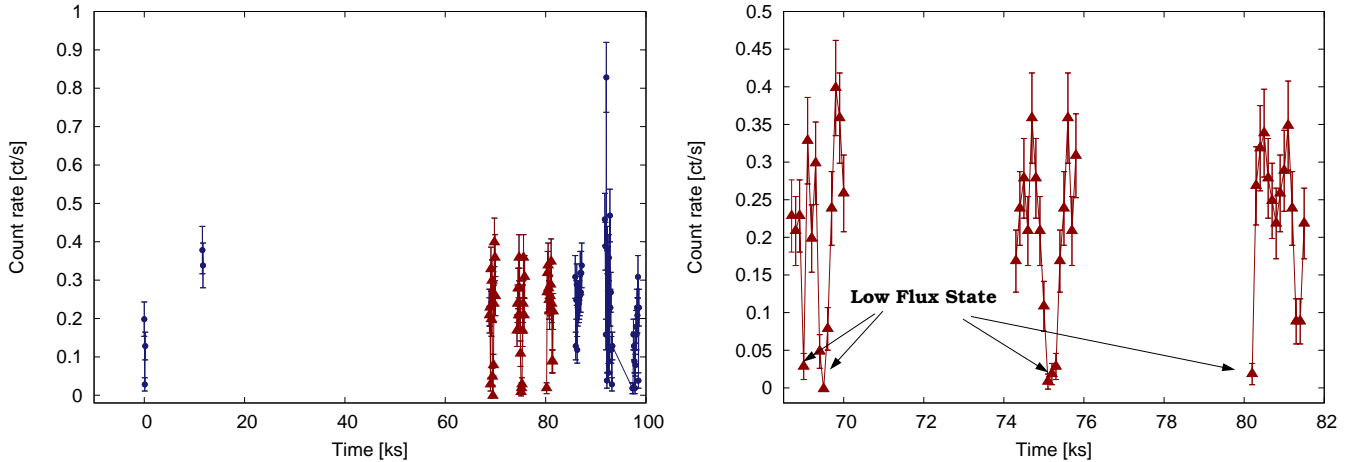}}
  \end{center}

  \caption{\textit{Left Panel:} Background-subtracted XRT count rate
    measured in the $0.5-10$\,keV band on 2013 October 18-19. Each data point is a 100-s
    average and the red triangles refer to the zoomed-in panel on the
    right. \textit{Right Panel:} Example of the X-ray flickering
    discussed in the text. Four episodes are highlighted in the plot,
    corresponding to X-ray fluxes compatible with the quiescent X-ray
    luminosity.}\label{fig1}

\end{figure*}

\subsection{X-Rays}

The October 18 \textit{Swift}/XRT observation (i.e., the first taken
after June 12) shows a dramatic, 20-fold increase in the count rate
(0.2 ct/s). The average flux remains approximately stable in all eight
exposures with a variation of less than a factor two in the average
fluxes.  We performed an X-ray spectral fit to the combined
${\sim}10$\,ks of data by using an absorbed power law model and
obtained a good fit ($\chi^2$/dof = 58/61) with a spectral index
$\Gamma = 1.69\pm 0.09$, a negligible absorption column
$N_H=3.8^{+2.0}_{-1.9}\times10^{20}\rm\,cm^{-2}$ (as found in
\citealt{arc10} and \citealt{bog11}) and a 0.5--10\,keV luminosity of
$2.5\times10^{33}\ergs$. The addition of a black-body component gives
unconstraining temperatures, but fixing the emission radius to 10 km
gives a 3-$\sigma$ upper limit of 0.127 keV and a luminosity of
$9\times10^{29}\ergs$.

Inspection of the light-curve on shorter timescales ($10-100$\,s)
reveals a surprising X-ray flickering with strong variability that
changes the flux by one order-of-magnitude. In some cases the flickering is observed to happen within
10\,s.  Shorter timescales cannot be investigated here because of the
limited count rate.  The flickering is visible several times in
the light-curve (see Figure~\ref{fig1}) and is observed
also when dividing the light-curve into two energy bands (a
$0.5-2.0$\,keV soft and a $2.0-10$\,keV hard band) with similar
magnitude. In one instance we observe a low-flux state followed by a
strong (100-fold) increase in flux, with a peak count rate of
${\sim}1$\,ct/s (Figure~1). We therefore infer an X-ray luminosity
variation from $10^{32}\ergs$ up to $10^{34}\ergs$ in less than 10\,s.

A power spectrum of the $0.5-10$\,keV light-curve shows no prominent
periodicity. 

\subsection{Optical/UV}

The UVOT images show that there is also a clear state change in the
UV/optical emission. On October 18, the UW1 counterpart is
very luminous with magnitude of $16.32\pm0.05$, which is 
${\sim}3.5$\,mag brighter than the previous UW1 observation on June
12. The U band shows also a bright counterpart with magnitude of $16.04\pm0.03$.

\begin{figure}[th!]
  \begin{center}
    \rotatebox{0}{\includegraphics[width=1.0\columnwidth]{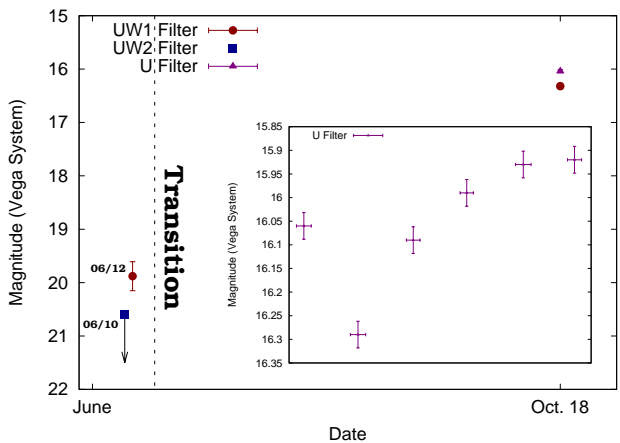}}
  \end{center}

  \caption{\textit{Swift}/UVOT observations in the UW1 (central
    $\lambda{\simeq}260$\,nm), UW2 (192.8\,nm) and U (346.5\,nm)
    bands. The source is not detected in UW2 on June 10 whereas it is
    marginally detected in UW1 on June 12. The error bars on the
    October 18/19 measurements are within the symbol size in the
    plot. Between June 15-23 the source turned off in radio
    \citep{sta13}. On October 18 the UW1 filter shows a clear increase
    in luminosity with the source detected also in the U band (purple
    triangle). The inset shows the six U-filter exposures with the
    horizontal bars that identify the length (${\sim}25$\,min
    per exposure).}\label{fig2}

\end{figure}

The six closely spaced exposures taken (October 19) show a U-band
variability of ${\sim}0.3$ mag on a timescale of ${\sim}1$ hr (see
Figure~\ref{fig2}) whereas no correlation is observed between the
average $0.5-10$\,keV X-ray luminosity and the U magnitude.

\subsection{Radio}

No radio emission was detected in the October observations with
10-$\sigma$ flux density upper limits of 0.8\,mJy, 0.2\,mJy and
0.1\,mJy for the full integration times of the 350\,MHz, 1380\,MHz,
and 5\,GHz observations, respectively.  These are all more than an
order-of-magnitude lower than the measured radio fluxes prior to this
current X-ray active phase (see~\citealt{sta13b, arc13}).  The
observations at 5\,GHz overlapped with two of the dips seen in the
\textit{Swift} data, (starting at MJD 56584.347264 \& MJD
56584.355933) but there is no evidence of an associated short burst of
pulsed radio emission. Furthermore, J1023 was previously detectable in
10-s intervals with Westerbork/Lovell at 350 and 1380\,MHz.  This also
gives confidence that the non-detection of the radio pulsar in short
sub-integrations is a significant change.

\section{Discussion}

We have reported the unambiguous X-ray and UV/optical signatures
of a state change in J1023, which coincide with its disappearance as
an observable radio MSP. This is the second time that such a state is
observed in J1023 and the first time that targeted X-ray data are
recorded in this state. The X-rays show a considerably softer spectrum
($\Gamma\simeq1.7$) than during the MSP state
($\Gamma\simeq 1$,~\citealt{arc10,bog11}) indicating a change in the
origin of the X-ray emission.

The large UV increase between June 12 and October 18 and the rapid
(${\sim}1$\,hr) changes in luminosity in the U band support the
presence of an accretion disk in the system.  \citet{bon02} reported
both slow (hours-to-days) and rapid (minutes-to-seconds) optical
flickering for J1023 in late 2000, during the previous active
phase. The flickering was observed on 10-s timescales, very similar to
what we now observe in X-rays.  
V band monitoring of the source~\citep{hal13} 
in its current state revealed a counterpart with magnitude between
16.37--16.94.  The similarity of the V, U and UW1 magnitudes is
compatible with the flatness of the continuum in a reprocessed
accretion disk ($F_{\nu}\propto \nu^{0.1}$).
\citet{hal13} also observed a double-peaked H$\alpha$ line, typical of
accretion disks. Such signatures were also identified by
\citet{bon02} in the 2001 active episode. The presence of an
accretion disk around J1023 does not, however, necessarily imply that
material is reaching the NS. Indeed, the total 0.5--10\,keV
luminosity remains low, with an average value of
2$\times10^{33}\ergs$, which is about $20\times$ higher than observed
during the MSP phase but still within the range of luminosities
of quiescent LMXBs~\citep{bil01}.

According to \citet{vanp94}, the absolute visual magnitude of LMXBs
correlates with the orbital period $P_{b}$ of the binary and its X-ray
luminosity. \citet{rus06} updated and revised this correlation,
separating black hole and NS LMXBs.  We therefore use the flux density
detected in the U band and rescaled to the distance of the pulsar to
place J1023 in the optical/X-ray diagram among the other known NS
LMXBs~\citep{rus06, rus07}.  We use the U band even if the original
correlation is calculated for the B,V,R,I and NIR bands because we
expect a negligible difference in reddening (which is ${\sim}0.1$\,mag
towards J1023) and a rather flat optical spectrum.

\begin{figure}[t!]
  \begin{center}
    \rotatebox{-90}{\includegraphics[width=0.7\columnwidth]{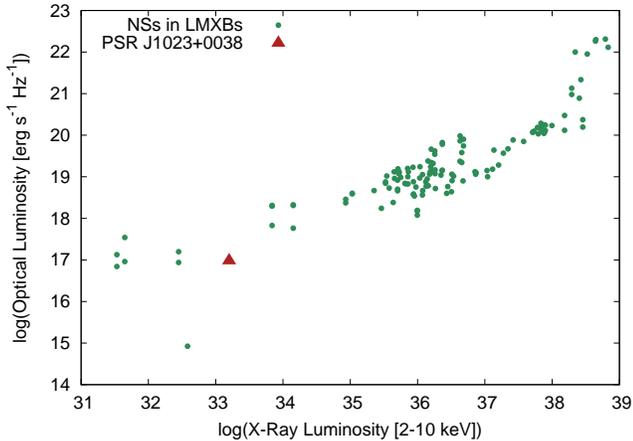}}
  \end{center}

  \caption{X-ray/optical correlation found in NS LMXBs with
    the addition of J1023. The correlation exists between the
    reprocessed light and the X-ray emission coming from the accretion
    process.}\label{fig3}

\end{figure}

Although there is significant scatter in the data points, J1023
fits into the X-ray/optical correlation (see Figure~\ref{fig3}),
further strengthening the idea that most of the optical and X-ray
luminosity comes from an accretion process and not from an intrabinary
shock or from the spin-down luminosity of a rotation-powered pulsar.

\citet{arc09} proposed that the low X-ray luminosity of the system
when an accretion disk is present is due to the onset of a propeller
stage. Indeed, given the spin-down-inferred $1.1\times10^8$\,G
magnetic field, Archibald et al. estimated that the required minimum
luminosity for accreting gas to overcome the magnetic centrifugal
barrier is $10^{37}\ergs$, well above the luminosities currently
reached by J1023.  The co-rotation radius $r_{c}$, where the Keplerian
orbital frequency of the matter equals the rotational frequency of the
NS (592\,Hz here), is located at about 24\,km from the NS center. The
onset of the propeller stage \citep{ill75} happens when $r_m>r_c$.  We
use the magnetospheric radius definition:
\begin{equation}
r_m=23.5 \xi \dot{M}_{-10}^{-2/7} M^{-1/7}_{1.4} R_{10}^{12/7}B_8^{4/7} \mbox{km}\label{mag}
\end{equation}
where $\xi=0.3-1$ is related to the structure of the inner disk, $B_8$
is the magnetic field normalized to $10^8$ G and $\dot{M}_{-10}$ is
the mass accretion rate in units of $10^{-10}\msun\rm\,yr^{-1}$.  The
condition $r_m>r_{c}$ will be met for a mass accretion rate of
approximately $10^{-10}\msun\rm\,yr^{-1}$, in line with the value
estimated by \citet{wan09b} for the previous 2001 active
episode. However, as noted by \citet{spr93} and \citet{rap04}, a
propeller regime where matter is expelled from the system needs $r_m
\gg r_c$. When this condition is not met, the matter will not be
expelled but will accumulate at the inner disk edge possibly leading
to a new disk structure, different from the standard Shakura \&
Sunyaev thin disk and known as a ``dead-disk'' \citep{siu77}. In
certain conditions, the dead-disk has the characteristic of having an
inner disk radius that remains close to $r_c$ even when the mass
accretion rate drops to zero (``trapped disk'', \citealt{dan10,dan12}).

The low flux states resemble the quasi-periodic dips observed in
several LMXBs (e.g.,~\citealt{dia06}), although we do not find any
strict correlation between their occurrence and a specific orbital
phase. The inclination of the orbit (${\sim}
46^{\circ}$~\citealt{bog11}) also strongly argues against the dips
interpretation.  If the rapid X-ray flickering is instead due to variation in
mass accretion rate at the inner disk edge, then for a standard
Shakura-Sunyaev disk, $r_m$ would move by a factor 2--3.  The
timescale for this drift is given by the viscous timescale:
\begin{eqnarray}
\tau_{\rm visc} & \sim & 3.5 \alpha^{-4/5}
\left[\frac{\dot{M}}{10^{-10}~{\rm M_{\odot}\,yr^{-1}}}\right]^{-3/10}
\left[\frac{M}{1~M_\odot}\right]^{1/4} \nonumber \\ && \times \left[\frac{\Delta\,R}{10~ {\rm
      km}}\right]^{5/4} s
\label{tvisc}
\end{eqnarray}
where $\alpha{\sim}0.1$ and $\Delta\,R$ represents the width of the
region involved.  To reach a timescale of 10--100\,s for
$\dot{M}{\sim}10^{-10}\rm\msun\,yr^{-1}$ we need to have an annulus of
no more than 10--100\,km.  Since the light cylinder radius is
$r_{lc}\simeq 80$ km, and if we assume that $r_m$ at the maximum
luminosity is located close to $r_{m}\simeq r_{c}\simeq 24$\,km, then
it is likely that $r_{m}$ does not lie far from $r_{lc}$ when the
system reaches the low flux state luminosity. In this case it is not
unreasonable to expect that the quenching of the radio emission would
stop and the pulsar would turn on again as an MSP.  Our radio search
for pulsations in the low flux states, however, found no detectable
radio pulses when the binary reaches the minimum X-ray flux level.  If
a trapped-disk is actually present, then the inner disk edge will
\textit{always} stay close to $r_{c}$ at about 24\,km even when
$\dot{M}\rightarrow 0$. In this case the quenching will always be
present during the active phase.

This radio non-detection is perhaps not surprising, since the ``radio
ejection'' mechanism originally suggested by \citet{shv70} is thought
to clear the disk from any system with a radio pulsar that has not
been quenched by material entering its light cylinder. However,
\citet{eks05} found that there is a region out to 2--3 light
cylinder radii where a disk can withstand the wind of an active
pulsar. If the disk around J1023 recedes into this regime,
the radio pulsar mechanism might be active some of the time.

Indeed, \citet{sta13} reported that since 2013 June 21 the gamma-ray
luminosity of J1023 has increased by a factor 5 with respect to the
average pre-transition luminosity. This is difficult to explain in an
active accretion state but it might indicate an enhancement
in the intra-binary shock emission, perhaps due to the pulsar wind
running into the inner edge of the disk or expelled gas colliding with
circumbinary material or the interstellar medium.

X-ray flickering, similar to that reported here, has also been
observed in the AMXP M28I \citep{pap13a,fer13}. This source is so far
the only AMXP known that has turned on as an active radio
MSP~\citep{pap13a}. During M28I's 2013 outburst, \textit{XMM-Newton}
observations showed a rapid flickering in the X-ray light-curve which
resembles what we observe in J1023, with the difference that the
flickering there happened at luminosities two/three
orders-of-magnitude higher than in J1023.  A power-spectrum showed red
noise but no specific periodicity.  The same source also showed fast
X-ray luminosity variations during quiescence, as seen with
\textit{Chandra} \citep{lin13}. Such variations span a factor of 7 in
luminosity (between ${\sim}10^{33}$ and $10^{34}\ergs$) on a timescale
of a few hundred seconds. The observation of more rapid timescales in
quiescence is however hampered by the large distance to M28I
(5.5\,kpc). \citet{lin13} reported that M28I becomes harder when going
from luminosities of $10^{34}\ergs$ down to $10^{32}\ergs$, which is
also observed in J1023.  When going towards luminosities above
$10^{34}\ergs$, the spectra again become harder, suggesting that some
new physical mechanism is at play.

A remaining open question is why M28I has gone through a relatively
bright outburst whereas J1023 shows only very faint luminosities (so
far at least). The higher spin rate of J1023 creates a faster rotating
magnetic field and thus a stronger magnetic centrifugal barrier.  
Compared with M28I, this likely makes it easier to achieve a propeller or dead-disk phase in J1023.
When J1023 returns to quiescence,
radio timing observations may be able to measure any spin-down or
spin-up that accumulates during this active phase, possibly
distinguishing between propeller and dead-disk models.

\acknowledgements{ We thank D. Russell for providing the optical/X-ray
  LMXB data of Fig.3. We thank the \textit{Swift} team for promptly scheduling
  the X-ray and UV observations. A.P. acknowledges support from the Netherlands
  Organization for Scientific Research (NWO) Vidi fellowship.
  A.~M.~A. and J.~W.~T.~H. acknowledge funding for this work from an
  NWO Vrije Competitie grant. The WSRT is operated by ASTRON with
  support from NWO.}


\begin{thebibliography}{25}
\expandafter\ifx\csname natexlab\endcsname\relax\def\natexlab#1{#1}\fi

\bibitem[{{Archibald} {et~al.}(2009){Archibald}, {Stairs}, {Ransom}, {Kaspi},
  {Kondratiev}, {Lorimer}, {McLaughlin}, {Boyles}, {Hessels}, {Lynch}, {van
  Leeuwen}, {Roberts}, {Jenet}, {Champion}, {Rosen}, {Barlow}, {Dunlap}, \&
  {Remillard}}]{arc09}
{Archibald}, A.~M., {Stairs}, I.~H., {Ransom}, S.~M., {et~al.} 2009, Science,
  324, 1411

\bibitem[{{Archibald} {et~al.}(2010){Archibald}, {Kaspi}, {Bogdanov},
  {Hessels}, {Stairs}, {Ransom}, \& {McLaughlin}}]{arc10}
{Archibald}, A.~M., {Kaspi}, V.~M., {Bogdanov}, S., {et~al.} 2010, \apj, 722,
  88

\bibitem[{{Archibald} {et~al.}(2013){Archibald}, {Kaspi}, {Hessels}, {Stappers},
  {Janssen} \& {Lyne}}]{arc13}
{Archibald}, A.~M., {Kaspi}, V.~M., {Hessels}, J.~W.~T., {et~al.} 2013,  arXiv:1311.5161

\bibitem[Bildsten 
\& Rutledge(2001)]{bil01} Bildsten, L., \& Rutledge, R.~E.\ 2001, The Neutron Star - Black Hole Connection, 245 


\bibitem[{{Bogdanov} {et~al.}(2011){Bogdanov}, {Archibald}, {Hessels}, {Kaspi},
  {Lorimer}, {McLaughlin}, {Ransom}, \& {Stairs}}]{bog11}
{Bogdanov}, S., {Archibald}, A.~M., {Hessels}, J.~W.~T., {et~al.} 2011, \apj,
  742, 97

\bibitem[{{Bond} {et~al.}(2002){Bond}, {White}, {Becker}, \& {O'Brien}}]{bon02}
{Bond}, H.~E., {White}, R.~L., {Becker}, R.~H., \& {O'Brien}, M.~S. 2002,
  \pasp, 114, 1359

\bibitem[{{D'Angelo} \& {Spruit}(2010)}]{dan10}
{D'Angelo}, C.~R., \& {Spruit}, H.~C. 2010, \mnras, 406, 1208

\bibitem[D'Angelo 
\& Spruit(2012)]{dan12} D'Angelo, C.~R., \& Spruit, H.~C.\ 2012, \mnras, 420, 416 

\bibitem[{{Deller} {et~al.}(2012){Deller}, {Archibald}, {Brisken},
  {Chatterjee}, {Janssen}, {Kaspi}, {Lorimer}, {Lyne}, {McLaughlin}, {Ransom},
  {Stairs}, \& {Stappers}}]{del12}
{Deller}, A.~T., {Archibald}, A.~M., {Brisken}, W.~F., {et~al.} 2012, \apjl,
  756, L25

\bibitem[D{\'{\i}}az Trigo et 
al.(2006)]{dia06} D{\'{\i}}az Trigo, M., Parmar, A.~N., Boirin, L., M{\'e}ndez, M., \& Kaastra, J.~S.\ 2006, \aap, 445, 179 

\bibitem[Ek{\c s}{\.I} \& Alpar(2005)]{eks05} Ek{\c s}{\.I}, K.~Y., \& Alpar, M.~A.\ 2005, \apj, 620, 390

\bibitem[Ferrigno et al.(2013)]{fer13} Ferrigno, C., Bozzo, 
E., Papitto, A., et al.\ 2013, arXiv:1310.7784 

\bibitem[{{Halpern} {et~al.}(2013)}]{hal13}
{Halpern}, J.~P., 2013 The Astronomer's Telegram, 5514, 1

\bibitem[Homer et al.(2006)]{hom06} Homer, L., Szkody, P., 
Chen, B., et al.\ 2006, \aj, 131, 562 

\bibitem[Karuppusamy et al.(2008)]{kar08} Karuppusamy, R., 
Stappers, B., \& van Straten, W.\ 2008, \pasp, 120, 191 


\bibitem[{{Kong} (2013)}]{kon13}
{Kong}, A.~K.~H., 2013 The Astronomer's Telegram, 5515, 1

\bibitem[{{Illarionov} \& {Sunyaev}(1975)}]{ill75}
{Illarionov}, A.~F., \& {Sunyaev}, R.~A. 1975, \aap, 39, 185

\bibitem[{{Linares} {et~al.}(2013)}]{lin13}
{Linares}, M.~A., {et al.} 2013 submitted to MNRAS


\bibitem[{{Papitto} {et~al.}(2013{\natexlab{a}}){Papitto}, {Ferrigno}, {Bozzo},
  {Rea}, {Pavan}, {Burderi}, {Burgay}, {Campana}, {di Salvo}, {Falanga},
  {Filipovi{\'c}}, {Freire}, {Hessels}, {Possenti}, {Ransom}, {Riggio},
  {Romano}, {Sarkissian}, {Stairs}, {Stella}, {Torres}, {Wieringa}, \&
  {Wong}}]{pap13a}
{Papitto}, A., {Ferrigno}, C., {Bozzo}, E., {et~al.} 2013{\natexlab{a}}, \nat,
  501, 517

\bibitem[{{Papitto} {et~al.}(2013{\natexlab{b}}){Papitto}, {Hessels}, {Burgay},
  {Ransom}, {Rea}, {Possenti}, {Stairs}, {Ferrigno}, \& {Bozz}}]{pap13b}
{Papitto}, A., {Hessels}, J.~W.~T., {Burgay}, M., {et~al.} 2013{\natexlab{b}},
  The Astronomer's Telegram, 5069, 1

\bibitem[{{Patruno} {et al.} (2013)}]{pat13b}
{Patruno}, A., {et al.}, 2013 The Astronomer's Telegram, 5516, 1

\bibitem[Patruno \& Watts(2012)]{pat12r} Patruno, A., \& Watts, A.~L.\ 2012, arXiv:1206.2727 

\bibitem[{{Rappaport} {et~al.}(2004){Rappaport}, {Fregeau}, \&
  {Spruit}}]{rap04}
{Rappaport}, S.~A., {Fregeau}, J.~M., \& {Spruit}, H. 2004, \apj, 606, 436

\bibitem[Roberts(2011)]{rob11} Roberts, M.~S.~E.\ 2011, 
American Institute of Physics Conference Series, 1357, 127 

\bibitem[{{Russell} {et~al.}(2006){Russell}, {Fender}, {Hynes}, {Brocksopp},
  {Homan}, {Jonker}, \& {Buxton}}]{rus06}
{Russell}, D.~M., {Fender}, R.~P., {Hynes}, R.~I., {et~al.} 2006, \mnras, 371,
  1334

\bibitem[{{Russell} {et~al.}(2007){Russell}, {Fender}, \& {Jonker}}]{rus07}
{Russell}, D.~M., {Fender}, R.~P., \& {Jonker}, P.~G. 2007, \mnras, 379, 1108

\bibitem[Shvartsman(1970)]{shv70} Shvartsman, V.~F.\ 1970,
\azh, 47, 660

\bibitem[{{Sunyaev} \& {Shakura}(1977)}]{siu77}
{Sunyaev}, R.~A., \& {Shakura}, N.~I. 1977, Pisma v Astronomicheskii Zhurnal,
  3, 262

\bibitem[{{Spruit} \& {Taam}(1993)}]{spr93}
{Spruit}, H.~C., \& {Taam}, R.~E. 1993, \apj, 402, 593

\bibitem[{{Stappers} {et~al.}(2013a)}]{sta13}
{Stappers}, B.~W., {et al.} 2013a The Astronomer's Telegram, 5513, 1

\bibitem[{{Stappers} {et~al.}(2013b)}]{sta13b}
{Stappers}, B.~W., {et al.} 2013b, submitted to ApJ

\bibitem[{{Tam} {et~al.}(2010){Tam}, {Hui}, {Huang}, {Kong}, {Takata}, {Lin},
  {Yang}, {Cheng}, \& {Taam}}]{tam10}
{Tam}, P.~H.~T., {Hui}, C.~Y., {Huang}, R.~H.~H., {et~al.} 2010, \apjl, 724,
  L207

\bibitem[{{Tauris}(2012)}]{tau12}
{Tauris}, T.~M. 2012, Science, 335, 561

\bibitem[Thorstensen 
\& Armstrong(2005)]{tho05} Thorstensen, J.~R., \& Armstrong, E.\ 2005, \aj, 130, 759 

\bibitem[{{van Paradijs} \& {McClintock}(1994)}]{vanp94}
{van Paradijs}, J., \& {McClintock}, J.~E. 1994, \aap, 290, 133

\bibitem[{{Wang} {et~al.}(2009){Wang}, {Archibald}, {Thorstensen}, {Kaspi},
  {Lorimer}, {Stairs}, \& {Ransom}}]{wan09b}
{Wang}, Z., {Archibald}, A.~M., {Thorstensen}, J.~R., {et~al.} 2009, \apj, 703,
  2017

\bibitem[{{Wijnands} \& {van der Klis}(1998)}]{wij98}
{Wijnands}, R., \& {van der Klis}, M. 1998, \nat, 394, 344

\end{thebibliography}

\end{document}